\documentclass[a4paper,11pt]{report}

\usepackage{graphicx}
\usepackage{latexsym}
\usepackage{amssymb}

\title{{\bf Portuguese Study Groups' Reports\\ \vspace*{1cm}
Report on\\[5mm]``\underline{Modelling Power Network}"\\[5mm]``State Estimation and Correction"}\\
\vspace*{8cm}{\normalsize Problem presented by \underline{INESC} at the\\ $86^{\rm th}$ European Study Group with Industry\\
$7^{\rm th}$--$11^{\rm th}$ May 2012\\
Instituto Superior de Engenharia do Porto\\
Portugal\\
}
\author{}}
\date{September, 2013}

\begin{document}
\maketitle
\thispagestyle{empty}
\vspace*{18cm}
\begin{tabular}{ll}
\textbf{Problem presented by:} & \begin{tabular}{l} \underline{Vladimiro Miranda} (\underline{INESC Porto})\end{tabular}\\
\textbf{Report prepared by:} & \begin{tabular}{l}\\
P. B\'arcia (barcia@novasbe.pt),\\ P. Castelo Ferreira (pedro.castelo.ferreira@gmail.com),\\ P. Freitas (freitas@cii.fc.ul.pt)
\end{tabular}\end{tabular}

\newpage
\begin{abstract}
\underline{Problem description}\\

A  power  network  (nodes,  branches)  is  regulated  by  flow  equations  based  on  the  First and Second Kirchhoff Laws.\\

\noindent {\bf LAW 1}: the net flow in a node of the network is zero: $\sum_j F_{ij}+\sum_j F_{ji}=0$. The  network  topology  is  a  graph  that  may  be  described  by  a  branch-node  incidence  matrix  $T$ (composed of elements with values -1, 1 or 0 only). Nodal injections are described by a vector $L$. The First Law may be translated into the matrix equation
\begin{center}
$T\,F=L\ .$\\
\end{center}

\noindent {\bf LAW 2}: the flow in a branch is proportional to the difference in potential P between its extreme nodes: $F_{ij}=b_{ij}(P_i-P_j)$. This may be globally translated into a matrix equation where $B$ is a diagonal matrix:
\begin{center}
$F=B\,T^t\,P$
\end{center}

The combination of the two Laws produces a well-known circuit equation
\begin{center}
$T\,B\,T^t\,P = L\ \ \mathrm{or}\ \ Y\,P = L$
\end{center}
where $Y$ is sometimes called a nodal-admittance matrix and $P$ is a vector of nodal potentials.\\

\noindent {\bf Question 1}

Admit that in a network with $n$ nodes and $m$ branches, one has available $k$ measurements, with $k > n$. These measurements may by on a mix of injections $L$, nodal potentials $P$ and branch flows $B$.

Admit that these measurements are contaminated with noise. Therefore, the measurements do not form a set compatible with the circuit equation or the Kirchhoff Laws.

Admit that this noise is Gaussian, and independent for each measurement. Admit that the variance is any case is small.

One wishes therefore to find a set of Potentials $\hat{P}$ that would minimize some reasonable definition of an error between the measurement vector and the vector of values ($F$, $L$ or $P$) that is compatible with the circuit equations.\\

\noindent {\bf Question 2}

Admit that some of the measurement errors are gross errors (much larger than the errors admitted previously), and that it is unknown where such gross errors occur. These may severely contaminate the estimation of $\hat{P}$.

Discover  which  measurements  contain  gross  errors  (instead  of  small  errors)  and  achieve  an estimation of $\hat{P}$ ignoring these gross errors.\\

\noindent {\bf Question 3}

Admit now that there are switches scattered in the network branches. They can assume a state of open ($S = 0$) or closed ($S = 1$). An open switch interrupts the branch flow and eliminates this branch from the network (namely, from matrix $T$).

Admit that there are measuring devices that report each switch status.

Assume that, beside the $k$ measurements of ($F$, $L$ or $P$), some switch status signals are missing – so, the network topology becomes unknown.

The challenge is double: to guess correctly the network topology and thus to estimate $\hat{P}$.

\end{abstract}

%\refstepcounter{chapter}
\setcounter{chapter}{3}
\setcounter{section}{0}
\setcounter{equation}{0}

\section{Introduction}

Let us define a generic power network with $N$ branches labeled by $i,k=1,\ldots,N$ and a maximum of $N(N-1)/2$ branches labeled by $ik$ with known complex branch impedances $z_{ik}$
\begin{equation}
z_{ik}=R_{ik}+j\,X_{ik}\ \ ,\ \ z_{ki}=z_{ik}
\end{equation}
where $R_{ik}=R_{ki}$ is the resistance and $X_{ik}=X_{ki}$ is the inductance for each branch. Further assuming the existence of switchers $s_{ik}$ for each branch which can be either on or off
\begin{equation}
s_{ik}=s_{ki}=\left\{\begin{array}{rcl}1&,& \mathrm{switcher\ is\ on}\ ,\\ 0&,& \mathrm{switcher\ is\ off}\ ,\end{array}\right. 
\end{equation}
the admittance matrix $Y$ for this network is symmetric and explicitly defined as
\begin{equation}
Y=\left[\begin{array}{cccc}\displaystyle y_{1}+\sum_{i\neq 1}\frac{s_{1i}}{z_{1i}}&\displaystyle-\frac{s_{12}}{z_{12}}&\ldots&\displaystyle-\frac{s_{1N}}{z_{1N}}\\
\displaystyle-\frac{s_{12}}{z_{12}}&\displaystyle y_{2}+\sum_{i\neq 2}\frac{s_{2i}}{z_{2i}}&\ldots&\displaystyle-\frac{s_{2N}}{z_{2N}}\\
\vdots&\vdots&\ddots&\vdots\\
\displaystyle-\frac{s_{1N}}{z_{1N}}&\displaystyle-\frac{s_{2N}}{z_{2N}}&\ldots&\displaystyle y_{N}+\sum_{i\neq N}\frac{s_{Ni}}{z_{Ni}}\\

\end{array}\right]
\end{equation}
where $y_{i}$ are the Earth admittances for each node.
Following the conventions of the problem we define:
\begin{itemize}
\item[$L_i$ :] injections at each node $i$, i.e the current intensities injected (if positive) or available (if negative);
\item[$P_i$ :] the potentials at each node $i$ measured with respect to some reference potential.
\end{itemize}
Given such definitions the First and Second Kirchoff law's are equivalent to the  matricial equation
\begin{equation}
L_i=\sum_{k=1}^N Y_{ik}\,P_j\ \ ,\ \ i=1,\ldots,N\ ,
\label{Law_gen}
\end{equation}
where the potentials and injections are generally complex quantities
\begin{equation}
\begin{array}{rcl}
\displaystyle P_i&=&\displaystyle P_{\mathrm{R},i}+jP_{\mathrm{I},i}=|P_i|(\cos\phi_{P,i}+j\sin\phi_{P,i})\ ,\\[5mm]
\displaystyle L_i&=&\displaystyle L_{\mathrm{R},i}+jL_{\mathrm{I},i}=|L_i|(\cos\phi_{L,i}+j\sin\phi_{L,i})\ .
\end{array}
\label{P_L_complex}
\end{equation}

For a specific given power network some of the branches will not be present and some of the existing branches will not have a physical switcher such that only a subset of the modeled switchers will actually be actionable. Hence to model a specific network it will be considered that
\begin{itemize}
\item[] $s_{ik}=0$ : for non existent branches;
\item[] $s_{ik}=1$ : for existing branches without switchers;
\item[] $s_{ik}=\left\{0,1\right\}$ : for existing branches with switchers.
\end{itemize}

For analysis and benchmarking purposes we are considering per unit values for all quantities. For instance choosing some reference impedance $z_{\mathrm{ref}}$ and potential $P_{\mathrm{ref}}$ such that the measured values for $P$ and $L$ must be scaled by the reference potential $P_{\mathrm{ref}}$ and reference injection $L_{\mathrm{ref}}=P_{\mathrm{ref}}/z_{\mathrm{ref}}$. Also we consider 2 distinct types of networks: DC networks and the DC approximation to AC networks. We do not explicitly work on fully AC networks as the analysis requires a much longer computational time.

\subsection{DC networks}

This is the simpler type of network employed for a preliminary testing of the techniques and methods of network analysis. Only the branches resistance is considered such that $X_{ij}=0$ and all quantities are real. Also the admittances to Earth at each node are considered null such that the matrix $Y$ is real and symmetric and the law~(\ref{Law_gen}) is explicitly given by
\begin{equation}
L_i=\sum_{k\neq i}\frac{s_{ik}}{R_{ik}} \left(P_i-P_k\right)
\label{Law_real}
\end{equation}

\subsection{DC approximation to AC networks}

This approximation is commonly employed in the analysis of AC power networks. It relies in the fact that for AC power lines the impedance is
much bigger than the resistance, the Earth admittances are negligible, the absolute values of the potentials $|P_i|$ are approximately constant at all nodes and the potential phases $\phi_P$ are small. Hence assuming the following simplifications
\begin{itemize}
\item[] $X_{ik}\gg R_{ik}$, for all branches $ik$;
\item[] $y_{i}\approx 0$, for all nodes $i$;
\item[] $|P_i|\approx 1$, for all nodes $i$ (in values per unit);
\item[] $\cos\phi_{P,i}\approx 1$, for all nodes $i$;
\item[] $\sin\phi_{P,i}\approx \phi_{P,i}$, for all nodes $i$,
\end{itemize} 
and decomposing the law~(\ref{Law_gen}) into real and imaginary parts we obtain that
\begin{equation}
L_{R,i}=\sum_{k\neq i}\frac{s_{ik}}{X_{ik}} \left(\phi_{P,i}-\phi_{P,k}\right)\ \ ,\ \ L_{I,i}\approx 0\ .
\label{Law_AC_DC}
\end{equation}

\subsection{AC networks}

When higher accuracy measurements are available and it is intended to estimate the potentials $P$ and injections $L$
with an higher accuracy the exact equations can be considered. In such case it can be considered a Cartesian decomposition into
the real and imaginary components of law~(\ref{Law_gen})
\begin{equation}
\begin{array}{rcl}
\displaystyle L_{R,i}&=&\displaystyle \sum_{k=1}^N \left(Y_{R,ik}P_{R,k}-Y_{I,ik}P_{I,k}\right)\ ,\\
\displaystyle L_{I,i}&=&\displaystyle \sum_{k=1}^N \left(Y_{I,ik}P_{R,k}+Y_{R,ik}P_{I,k}\right)\ ,
\end{array}
\label{Complex_Cartesian}
\end{equation}
where $Y_{R,ik}$ and $Y_{I,ik}$ stand for the real and imaginary components of the complex matrix entries $Y_{ij}$.

In the analysis of power networks it is often considered the Euler form such that the several quantities are represented
by their absolute value and phase. Considering a decomposition of the complex matrix entries $Y_{ik}=|Y_{ik}|(\cos\phi_{Y,ik}+j\sin\phi_{Y,ik})$ and the decomposition~(\ref{P_L_complex}) for the $P$'s and $L$'s, the law~(\ref{Law_gen}) is expressed as
\begin{equation}
\begin{array}{rcl}
\displaystyle |L_i|&=&\displaystyle \sum_{k=1}^N|Y_{ik}||P_k|\frac{\cos\left(\phi_{Y,ik}+\phi_{P,k}\right)}{\cos(\phi_{L,i})}\ ,\\
\displaystyle \tan(\phi_{L,i})&=&\displaystyle \sum_{k=1}^N\tan(\phi_{Y,ik}+\phi_{P,k})\ .
\end{array}
\label{Complex_Euler}
\end{equation}

The Cartesian decomposition~(\ref{Complex_Cartesian}) has the advantage of representing the law~(\ref{Law_gen}) by a linear expression as opposed to the Euler form~(\ref{Complex_Euler}). Hence, as long as non-linear effects on networks are negligible, the Cartesian decomposition simplifies
the technical formulation and analysis of the network equations, however the estimative errors is lower when considering the Euler decomposition
than the Cartesian decomposition.

\section{State Estimation for known network topologies}

Generally the estimation of a network state is computed employing a weighted least square (WLS) method.
Given a set of $m$ measurements $A_i$ with measurement errors $\epsilon_{A,i}$ and a set of laws $A_0(\mathbf{B})$ depending on
the $N_B$ network parameters $B_j$, each measurement is expressed as
\begin{equation}
A_i=A_0(\mathbf{B})+\epsilon_{A,i}\ .
\end{equation}
Assuming that the measurement errors $\epsilon_{A,i}$ have null mean and a variance $\sigma_i^2$, the standard WLS minimization method relies on the definition of a quadratic objective function $J(\mathbf{\hat{B}})$ to be minimized with respect to the quantities to estimate $\hat{B}_j$
\begin{equation}
J(\mathbf{\hat{B}})=\sum_{i=1}^{m>N}\frac{\left(A_i-A_{0,i}(\mathbf{\hat{B}})\right)\left(A_i-A_{0,i}(\mathbf{\hat{B}})\right)^*}{\sigma_i^2}\ ,
\end{equation}
such that the solution of the system of $N_B$ equations
\begin{equation}
\frac{dJ}{d\hat{B}_j}=0\ ,\ j=1,\ldots,N_B\ ,
\end{equation}
constitutes the state estimate obtained from the $m$ measurements for a network with a state defined by $N_B$ parameters.
If any set of $N_g$ constraints $g_k(\mathbf{B})=0$ must be considered, these may be included in the quadratic objective function through the Lagrange multiplier method. Defining
\begin{equation}
J_g(\mathbf{\hat{B}})=J(\mathbf{\hat{B}})+\sum_{k}\lambda_k\,g_k(\mathbf{B})\ ,
\end{equation}
the network state estimation is the solution to the system of $N_B+N_g$ equation 
\begin{equation}
\left\{\begin{array}{rcl}\displaystyle\frac{dJ_c}{d\hat{B}_j}&=&0\ ,\ j=1,\ldots,N_B\ ,\\[5mm]
\displaystyle\frac{dJ_c}{d\hat{\lambda}_k}&=&0\ ,\ k=1,\ldots,N_g\ ,\end{array}\right.
\end{equation}

In the following we are considering only simultaneously measurements of nodal injections $L_i$ and potentials $P_i$, when some delay
between the actual measurements and the recording of its values exist it may be considered a synchronization which
performs the measurement in advance of the network analysis accounting for such delay. For a network with $N$ nodes we may have a maximum of $2N$ measurements, more generally some of the measurements can be absent such that we define the quantities $\tilde{L}_i$ and $\tilde{P}_i$ which
coincide with the measured quantities $L_i$ and $P_i$ when available or, otherwise, coincide with the quantities to estimate $\hat{L}_i$
and $\hat{P}_i$
\begin{equation}
\begin{array}{rcl}
\tilde{L}_i&=&\left\{\begin{array}{lcl}L_i&,&\mathrm{if\ }\exists_{\mathrm{measurement}}\ ,\\\hat{L}_i&,&\mathrm{if\ }\nexists_{\mathrm{measurement}}\ ,\end{array}\right.\\[6mm]
\tilde{P}_i&=&\left\{\begin{array}{lcl}P_i&,&\mathrm{if\ }\exists_{\mathrm{measurement}}\ ,\\\hat{P}_i&,&\mathrm{if\ }\nexists_{\mathrm{measurement}}\ .\end{array}\right.\\[6mm]
\end{array}
\end{equation}
If a branch flow $F_{ij}$ measurement or a branch power flow $S_{ij}$ exist can be included in the quadratic objective function by
considering the following constraints
\begin{equation}
\begin{array}{rcl}
P_i-P_j-z_{ij}F_{ij}&=&0\ ,\\[5mm]
(P_i-P_j)(P_i-P_j)^*-z_{ij}S_{ij}^*&=&0\ .
\end{array}
\end{equation}

Next we will test several definitions for the quadratic objective function and carry a statistical benchmark for the several network types discussed in the introduction. The most standard definition for this function is
\begin{equation}
\begin{array}{rcl}
J_1(\hat{P}_j,\tilde{L}_j=\hat{L}_j)&=&\displaystyle\sum_{i=1}^{N}\frac{\displaystyle\left(\tilde{P}_i-\hat{P}_i\right)\left(\tilde{P}_i-\hat{P}_i\right)^*}{\displaystyle\sigma_{P,i}^2}\\[5mm]
&&\hfill+\frac{\displaystyle\left(\tilde{L}_i-\sum_{j=1}^NY_{ij}\hat{P}_j\right)\left(\tilde{L}_i-\sum_{j=1}^NY_{ij}\hat{P}_j\right)^*}{\displaystyle\Sigma_i^2}\ .
\end{array}
\label{J_classic_1}
\end{equation}
We note that the specific expression for the weight $\Sigma_i$ does influence the error of the estimates for the potentials and injections with respect to the actual values. We will carry a preliminary analysis for several possible definitions of this weight later on.

The minimization of~(\ref{J_classic_1}) is performed with respect to the estimated quantities $\hat{P}_i$ and to the injections for which measurements do not exist $\tilde{L}_i=\hat{L}_i$. The most simple method to estimate the injections is to apply the law~(\ref{Law_gen}) to the estimated potentials
\begin{equation}
\hat{L}_i=\sum_{j=1}^NY_{ij}\hat{P}_j\ ,
\label{J_classic_10}
\end{equation}
such that these estimates for the injections $\hat{L}_i$ and the estimates for the potentials $\hat{P}_i$ is obtained from~(\ref{J_classic_1})
define de network state.

Alternatively it may be defined a quadratic objective function independent of the estimated values for the potentials $\hat{P}_i$ computed
from~(\ref{J_classic_1})
\begin{equation}
\begin{array}{rcl}
J_{2(1)}(\tilde{P}_j=\hat{P}_j,\hat{L}_j)&=&\displaystyle\sum_{i=1}^{N}\frac{\displaystyle\left(\tilde{L}_i-\hat{L}_i\right)\left(\tilde{L}_i-\hat{L}_i\right)^*}{\displaystyle\sigma_{L,i}^2}\\[5mm]
&&\hfill+\frac{\displaystyle\left(\hat{L}_i-\sum_{j=1}^NY_{ij}\tilde{P}_j\right)\left(\hat{L}_i-\sum_{j=1}^NY_{ij}\tilde{P}_j\right)^*}{\displaystyle\Sigma_i^2}\ .
\end{array}
\label{J_classic_11}
\end{equation}
Hence, minimizing this function we obtain a distinct estimate for the injections $\hat{L}_i$ which, together with the estimate for the potentials $\hat{P}_i$ obtained from~(\ref{J_classic_1}) defines the network state.

In addition the function~(\ref{J_classic_1}) can be minimized for the nodal potentials $\hat{P}_i$ simultaneously with the minimization, for the nodal injections $\hat{L}_i$, of the quadratic objective function
\begin{equation}
\begin{array}{rcl}
J_{2(2)}(\hat{L}_j)&=&\displaystyle\sum_{i=1}^{N}\frac{\displaystyle\left(\tilde{L}_i-\hat{L}_i\right)\left(\tilde{L}_i-\hat{L}_i\right)^*}{\displaystyle\sigma_{L,i}^2}\\[5mm]
&&\hfill+\frac{\displaystyle\left(\hat{L}_i-\sum_{j=1}^NY_{ij}\hat{P}_j\right)\left(\hat{L}_i-\sum_{j=1}^NY_{ij}\hat{P}_j\right)^*}{\displaystyle\Sigma_i^2}\ .
\end{array}
\label{J_classic_12}
\end{equation}
Hence solving the system of $N+N$ equations obtained from the simultaneously minimization of functions~(\ref{J_classic_1}) and~(\ref{J_classic_12}) we obtain an estimate for the network state.

Yet another possibility is to define a quadratic objective function that depends both on the estimate for potentials $\hat{P}_j$ and estimate for the injections $\hat{L}_j$
\begin{equation}
\begin{array}{rcl}
J_{3}(\hat{P}_j,\hat{L}_j)&=&\displaystyle\sum_{i=1}^{N}\frac{\displaystyle\left(\tilde{P}_i-\hat{P}_i\right)\left(\tilde{P}_i-\hat{P}_i\right)^*}{\displaystyle\sigma_{P,i}^2}+\frac{\displaystyle\left(\tilde{L}_i-\hat{L}_i\right)\left(\tilde{L}_i-\hat{L}_i\right)^*}{\displaystyle\sigma_{L,i}^2}\\[5mm]
&&\hfill+\frac{\displaystyle\left(\hat{L}_i-\sum_{j=1}^NY_{ij}\hat{P}_j\right)\left(\hat{L}_i-\sum_{j=1}^NY_{ij}\hat{P}_j\right)^*}{\displaystyle\Sigma_i^2}\ .
\end{array}
\label{J_classic_3}
\end{equation}
Minimizing this function with respect to both the potentials $\hat{P}_i$ and injections $\hat{L}_i$ we obtain an estimate for the network state.

Next we carry a numerical statistical benchmark of the four distinct set of quadratic objective function for the several network types.
In our analysis the exact values of the potentials and injections are $P_{0,i}$ and $L_{0,i}$  and the errors for the measurements $P_i$ and $L_i$ are assumed to be given as a percentage $\epsilon_\%$ of the actual values for $P_{0,i}$ and $L_{0,i}$ with a statistical Gaussian distribution of null mean and standard deviation $\sigma_\%$
\begin{equation}
\begin{array}{rcl}
\epsilon_\%&\sim&\mathcal{N}(0,\sigma_\%)\ ,\\[5mm]
\epsilon_{P,i}&=&P_{0,i}\epsilon_\%\ \Rightarrow\ \epsilon_{P,i}\sim \mathcal{N}(0,\sigma_{P,i})\ ,\\[5mm]
\epsilon_{L,i}&=&L_{0,i}\epsilon_\%\ \Rightarrow\ \epsilon_{L,i}\sim \mathcal{N}(0,\sigma_{L,i})\ ,\\[5mm]
\sigma_{P,i}&=&|P_{0,i}|\sigma_\%\ ,\\[5mm]
\sigma_{L,i}&=&|L_{0,i}|\sigma_\%\ .
\end{array}
\end{equation}    
These standard deviations correspond either to the instrumentation accuracy, when known, or can be directly computed from
measurement data assuming that the mean is null.

We note that the amount of improvement of the estimate errors with respect to the measurement error for all these
estimate procedures does depend in several of the network parameters and topology as well as on the per unit reference
quantities employed. Hence there is no unique choice for a better method or parameter scaling that can be universally applied
to all existing power networks. For exemplification purposes, in the following we are carrying a benchmark analysis based on a set of randomly generated networks and measurement errors. For a known network, a dedicated benchmark must be carried allowing for a significant
improvement of the estimate errors.

We are considering per unit values of the several quantities and choosing the reference values for $P_{\mathrm{ref}}$ such that $<|P_0|>\sim 1$ and $<\phi_{P,0}>\sim 0$ and will analyze the possible improvement on the estimate error with respect to the actual values of the potentials, injections and electric power for a network of $N$ nodes depending on the network type, the quadratic objective function and parameters:
\begin{itemize} 
\item network type: DC, DC approximation to AC or AC networks;
\item quadratic objective function: $J_1$, $J_{2(1)}$, $J_{2(2)}$ or $J_3$;
\item the definition of $\Sigma_i$;
\item the average value of impedances $<z>$: set by the reference impedance $z_{\mathrm{ref}}$;
\item the network connectivity $p$: the average connections per node;
\item the number of available measurements: $N<m\leq 2N$.
\end{itemize}

We split this analysis into the three network types in the following subsections and consider as the reference approach
that the weight $\Sigma_i$ is
\begin{equation}
\Sigma_i^2=\sigma_{L,i}^2\ ,
\label{Sigma_cl}
\end{equation}
the average impedance of $<z>$ is
\begin{equation}
<z>\sim 100\ .
\label{avez_cl}
\end{equation}
This value can be changed by choosing a distinct reference impedance $z_{\mathrm{ref}}$ which is equivalent
to an overall scaling of the matrix $Y$. For each of the network types we carry the benchmark of the several parameters by considering
random networks of 20 nodes and compute the average of the percentage of the estimate error with respect to the measurement errors for
a sample of random measurements for a random set of networks. We consider the following expressions for
$\Sigma_i$
\begin{equation}
\begin{array}{l}
\Sigma_{L,n,i}^2=\sigma_{L,i}^n\ \ , \ \ \Sigma_{P,n,i}^2=\alpha\left(\sigma_{P,i}\sigma_{P,i}\right)^{n/2}\ ,\\
\Sigma_{PL,n,i}^2=\left(\sigma_{L,i}^2+\sum_kY_{ik}^2\sigma_{P,k}^2\right)^{n/2}\ \ ,\ \ n\in\mathbb{N}\ ,
\end{array}
\label{Sigma_n}
\end{equation}
where for numerical stability we are considering $n=0,1,2,3,4,5,6$ to be an integer and $\alpha$ is a dimensionfull constant of value of unity that ensure that the functions $J$ are dimensionless. As for the average value of the impedances we analyze it in the range $<z>\in[0.1,500]$. We also note that the definition of $\Sigma_i$ does affect the hypotheses testing and statistical significance analysis of the estimate as it explicitly modifies the variance of the quadratic objective functions. We do not discuss such analysis here.
All quantities are generated randomly including the network topology and in the following analysis we are considering
\begin{equation}
\sigma_\%=0.1\ ,
\label{sigma_per}
\end{equation}
for the impedances
\begin{equation}
z\in z_{\mathrm{ref}}\times[0.5,1.5]\ ,
\label{z_ave}
\end{equation}
and for the number of branches on the network
\begin{equation}
n_{branch} = \frac{p}{N-1}\ ,
\label{n_branch}
\end{equation}
where $p$ is the average number of connections per node. Within each analysis the actual potentials at each node are kept fixed, while the topology and measurement errors is randomly varying and we will explicitly analyze the average percentage of the estimation error with respect to
the measurement error, i.e. for a given nodal quantity $A_i$
\begin{equation}
\left<\%\ \mathrm{error}_A\right>=100\%\,\left<\sqrt{\frac{(A_0-\hat{A})\cdot (A_0-\hat{A})}{\epsilon_{A}\cdot\epsilon_{A}}}\right>\ ,
\end{equation}
where '$\cdot$' represents a vectorial product, $A_{0,i}$, $\hat{A}_i$ and $\epsilon_{A,i}$ are the actual values, the estimated values and
the measurement errors, respectively, for the quantity $A$ at each node $i$ and the average is taken over successive measurements.

We will further include the estimative for nodal power flow, however we note that the procedures
described here are not adequate to estimate these quantities, as the quadratic objective functions do not explicitly include a minimization
for such quantities as have been written with the objective of minimizing only the $P$'s and $L$'s.

\subsection{DC networks}

\begin{figure}
\begin{center}
\includegraphics[width=130mm]{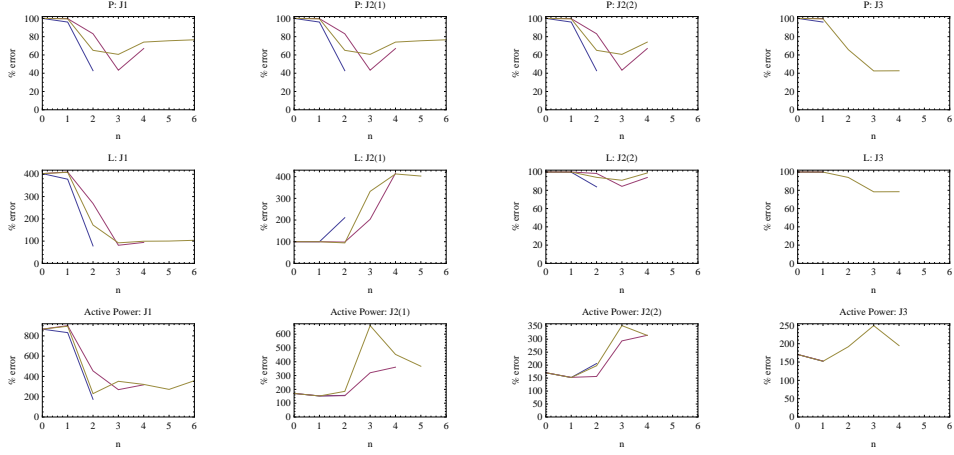}
\caption{Error percentage for the estimates for the potentials $P$, injections $L$ and active power, for $<z>\sim 100$ for DC networks when $m=2N$ measurements are available for $N=20$ nodes and a sampling of
50 random networks of connectivity $p=3$ and 100 measurements for each network; Blue: $\Sigma_{L,n,i}$, Magenta: $\Sigma_{P,n,i}$, Brown: $\Sigma_{PL,n,i}$. \label{fig.weigth_real_1_100}}
\end{center}
\end{figure}

\begin{figure}
\begin{center}
\includegraphics[width=130mm]{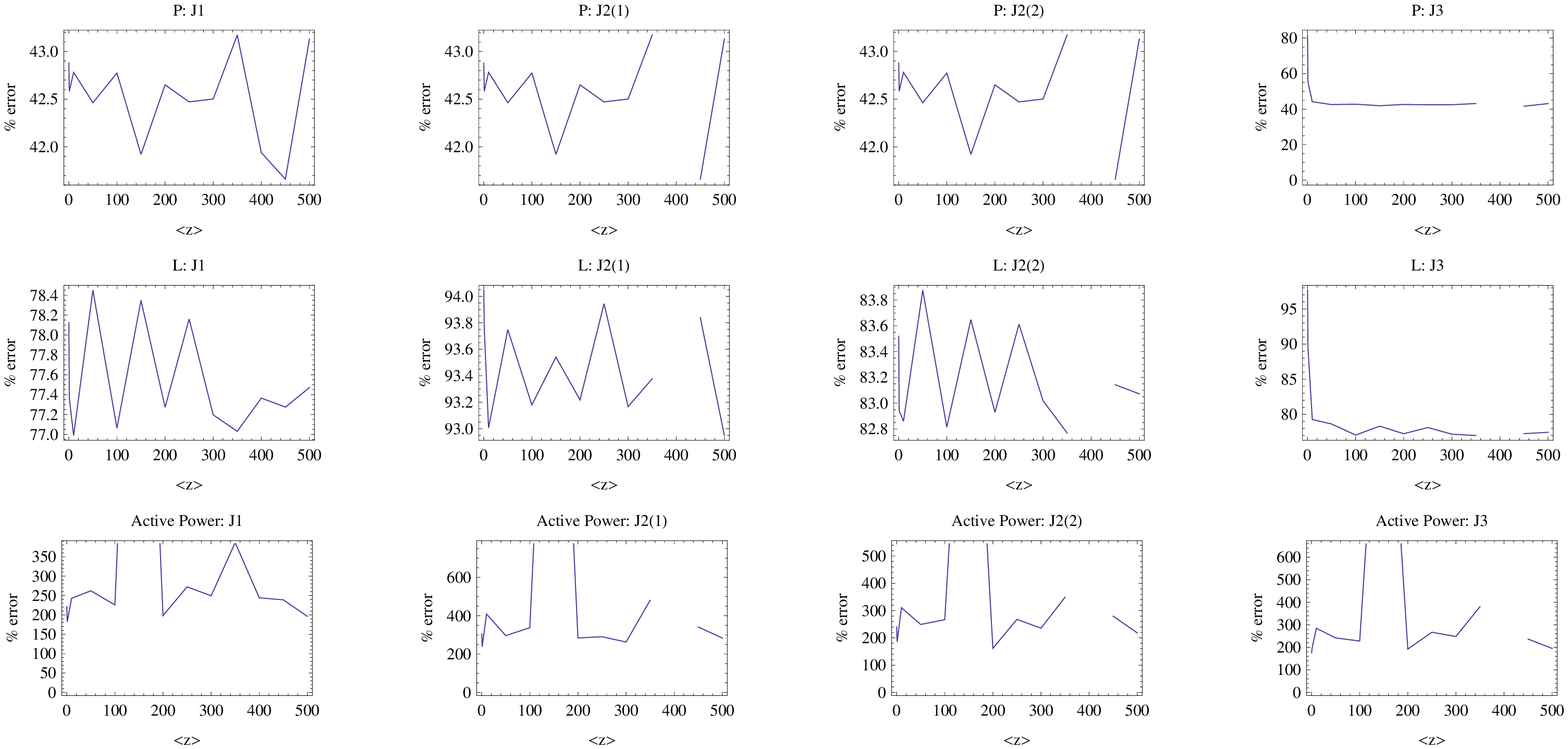}
\caption{Running with $<z>$ of percentage of estimate errors with respect to measurement errors for DC networks when $m=2N$ measurements are available for $N=20$ and a sampling of 50 random networks of connectivity $p=3$ and 100 measurements. \label{fig.stat_real_1_delZ}}
\end{center}
\end{figure}

For DC networks let us fix the average value for the impedances $<z>\sim 100$~(\ref{avez_cl}) and a connectivity of $p=3$
and consider the several possible quadratic objective functions and the several definitions for $\Sigma_i$ suggested in equation~(\ref{Sigma_n})
when $m=2N$ measurements are available. The average estimate error percentage with respect to the measurement error is plotted in figure~\ref{fig.weigth_real_1_100} for the several $J$'s as a function of $n$ for a sampling of 50 random networks and 100 measurements.
From direct inspection of these results the weights that allow for the lower estimate errors either for the potentials $P$ or the injections $L$
can be chosen, following the original problem posed. Hence the choice for each of the quadratic
objective functions are
\begin{equation}
\begin{array}{rl}
J_1:&\Sigma^2_i=\Sigma_{L,2,i}=\sigma_{L,i}^2\ (n=2)\ ,\\[5mm]
J_{2(1)}:&\Sigma^2_i=\Sigma^2_{PL,2,i}=\sigma_{L,i}^2+\sum_k Y_{ik}^2\sigma_{P,k}^2\ (n=2)\ ,\\[5mm]
J_{2(2)}:&\Sigma^2_i=\Sigma_{L,2,i}=\sigma_{L,i}^2\ (n=2)\ ,\\[5mm]
J_{3}:&\Sigma^2_i=\Sigma_{PL,3,i}=\left(\sigma_{L,i}^2+\sum_k Y_{ik}^2\sigma_{P,k}^2\right)^{3/2}\ (n=3)\ .\\[5mm]
\end{array}
\end{equation}
For these choices we plot in figure~\ref{fig.stat_real_1_delZ} the estimate errors dependence on the average value of the impedances $<z>$.
As there are no significant changes on the estimate errors with the value of $<z>$ in the neighborhood of $<z>\sim 100$ we keep
working with this value. We further note that for $J_2(1)$, $J_{2(2)}$ and $J_3$ for values above $<z>\geq 350$ not always exist solutions
for the minimizing equations.

\begin{figure}
\begin{center}
\includegraphics[width=130mm]{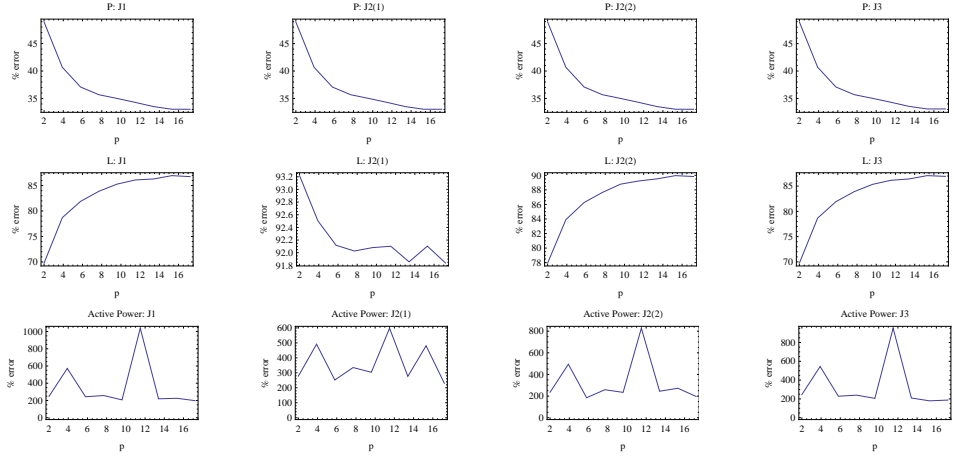}
\caption{Running with network connectivity $p$ of percentage of estimate errors with respect to measurement errors for DC networks when $m=2N$ measurements are available for $N=20$ nodes with $<z>\sim 100$ and a sampling of 50 random networks and 100 measurements. \label{fig.stat_real_1_100_P}}
\end{center}
\end{figure}

With respect to the network connectivity we plot the dependence of the estimate errors as a function of $p$~(\ref{n_branch}) in figure~\ref{fig.stat_real_1_100_P} for the several functions $J$'s and the above choices.

\begin{figure}
\begin{center}
\includegraphics[width=130mm]{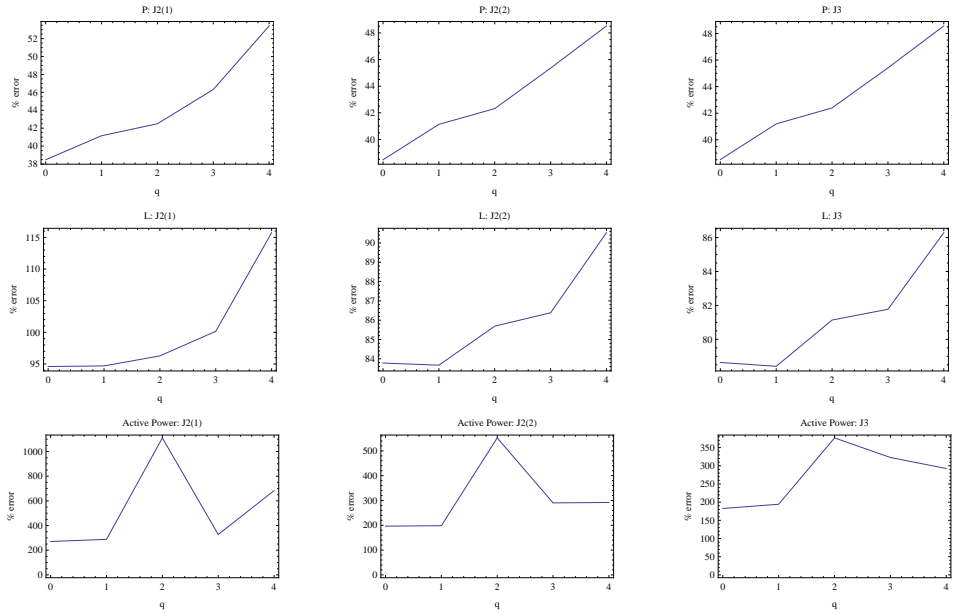}
\caption{Running with measurement availability $m=2N-q$ of percentage of estimate errors with respect to measurement errors for DC networks for $N=20$ nodes with $<z>\sim 100$ and $p=3$ and a sampling of (5)50 random networks and (10)100 measurements; Blue: $J_1$, Magenta: $J_{2(1)}$, Brown: $J_{2(2)}$, Green: $J_{3}$. \label{fig.stat_real_1_100_m}}
\end{center}
\end{figure}

Finally when only $m=2N-q$ measurements are available such that $N<m<2N$, we plot the dependence of the estimate errors as a function of $q$ in figure~\ref{fig.stat_real_1_100_m} for the several functions $J$'s.
We note that depending on the specific network being analyzed, for $q>6$, no exact solutions exist that minimize the functions $J$. Employing a numerical solver is possible to obtain convergent solutions up to $q=12$ within a given accuracy, however for $q>12$, generally it no convergent solution exist. In the particular case of $J_1$ only for $q=0$ and $q=1$ exist exact solutions that minimize it. Hence we conclude that the best minimizing function for DC networks of $N=20$ nodes which allows for less than $2N$ available measurements is either $J_3$ or $J_{2(2)}$ with the weights $\Sigma_i^2=\Sigma_{PL,3,i}$ and $\Sigma^2_i=\Sigma_{L,2,i}$, respectively.

\clearpage

\subsection{DC approximation to AC networks}

\begin{figure}
\begin{center}
\includegraphics[width=130mm]{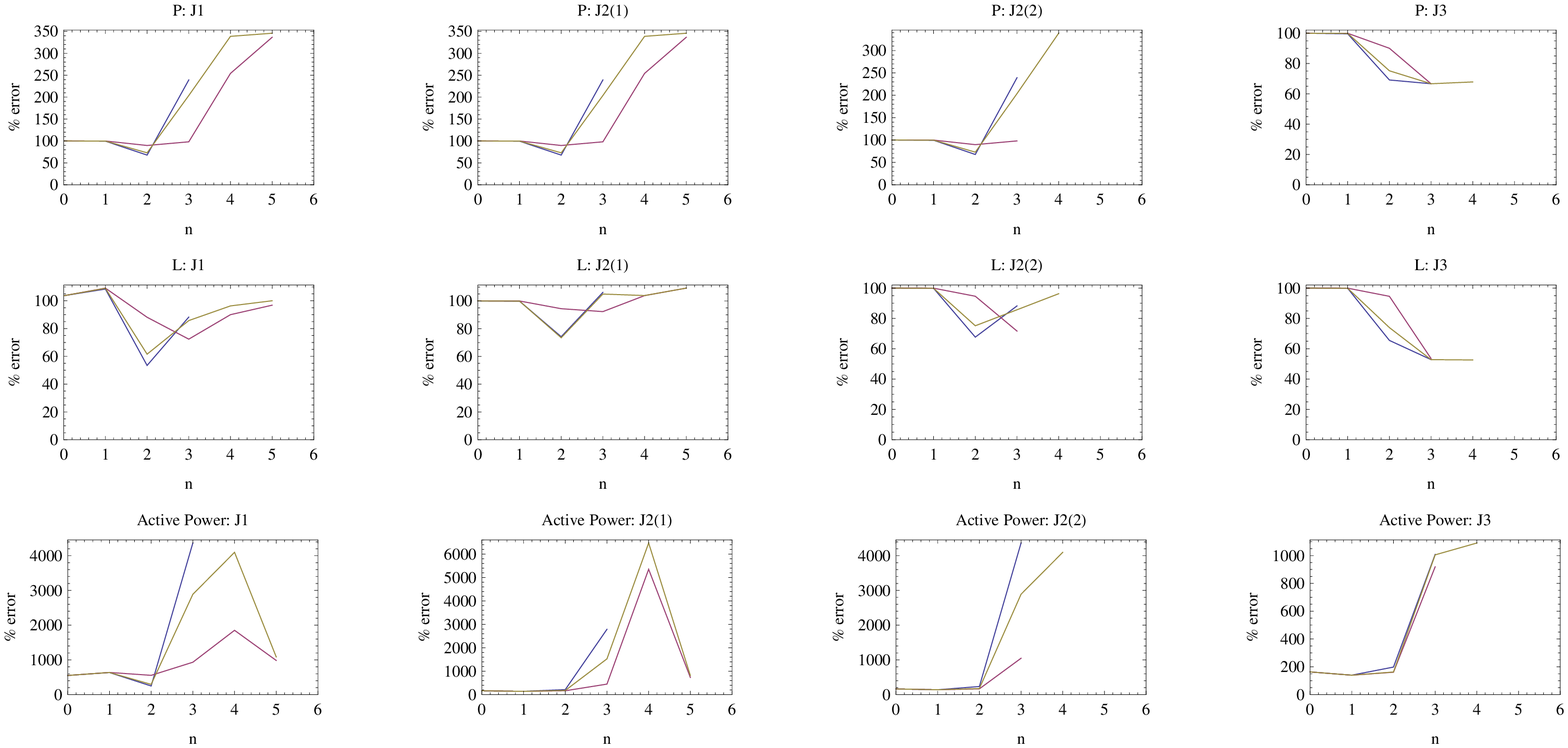}
\caption{Error percentage for the estimates for the potentials $P$, injections $L$ and active power, for $<z>\sim 100$ for DC approximation to AC networks when $m=2N$ measurements are available for $N=20$ and a sampling of 50 random networks of connectivity $p=3$ and 100 measurements; Blue: $\Sigma_{L,n,i}$, Mangenta: $\Sigma_{P,n,i}$, Brown: $\Sigma_{PL,n,i}$. \label{fig.weigth_real_0_100}}
\end{center}
\end{figure}

For the DC approximation to AC networks, fixing the impedances average value $<z>\sim 100$~(\ref{avez_cl}) and a connectivity of $p=3$
and consider the several possible quadratic objective functions and the several definitions for $\Sigma_i$ suggested in equation~(\ref{Sigma_n})
when $m=2N$ measurements are available. The average estimate error percentage with respect to the measurement error is plotted in figure~\ref{fig.weigth_real_0_100} for the several $J$'s as a function of $n$ for a sampling of 50 random networks and 100 measurements for each network.
Again, employing the criteria of minimization of the $P$'s and $L$'s the choice for each of the quadratic
objective functions are
\begin{equation}
\begin{array}{rl}
J_1:&\Sigma^2_i=\Sigma_{L,2,i}=\sigma_{L,i}^2\ (n=2)\ ,\\[5mm]
J_{2(1)}:&\Sigma^2_i=\Sigma_{L,2,i}=\sigma_{L,i}^2\ (n=2)\ ,\\[5mm]
J_{2(2)}:&\Sigma^2_i=\Sigma_{L,2,i}=\sigma_{L,i}^2\ (n=2)\ ,\\[5mm]
J_{3}:&\Sigma^2_i=\Sigma_{P,1,i}=\left(\sigma_{L,i}\sigma_{P,i}\right)^{3/2}\ (n=3)\ .\\[5mm]
\end{array}
\end{equation}

\begin{figure}
\begin{center}
\includegraphics[width=130mm]{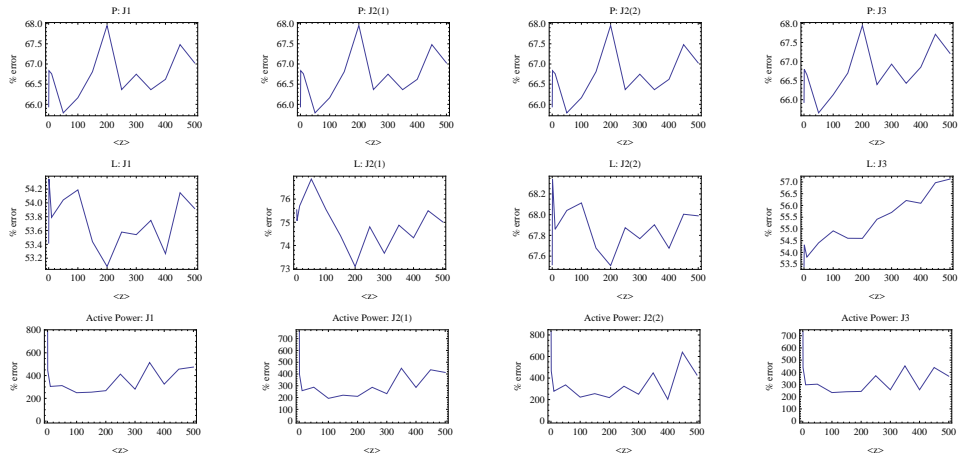}
\caption{Running with $<z>$ of percentage of estimate errors with respect to measurement errors for DC approximation to AC networks when $m=2N$ measurements are available for $N=20$ and a sampling of 50 random networks of connectivity $p=3$ and 100 measurements. \label{fig.stat_real_0_delZ}}
\end{center}
\end{figure}

For these choices we plot in figure~\ref{fig.stat_real_0_delZ} the dependence of the estimate errors as a function of
the average value of the impedances $<z>$
Again, there is no significant change on the estimative improvement with respect to the measured quantities for values of $<z>$ in the neighborhood of $<z>\sim 100$, hence we proceed with this value.

\begin{figure}
\begin{center}
\includegraphics[width=130mm]{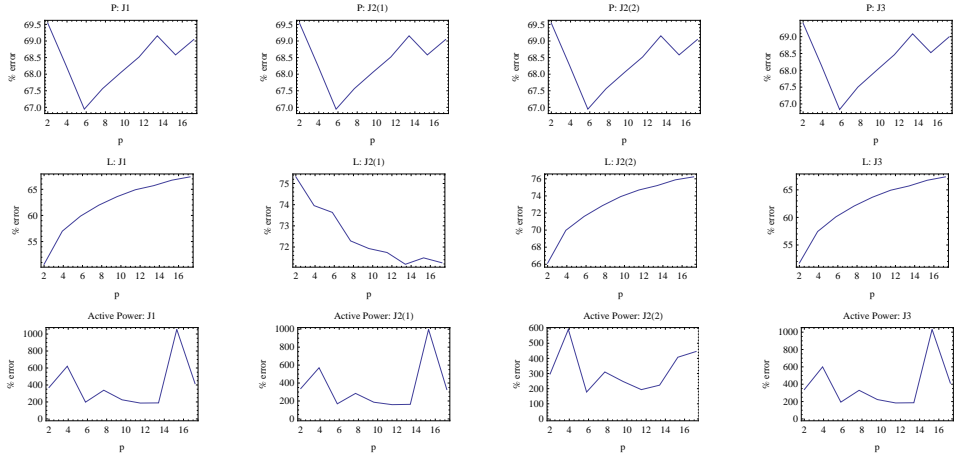}
\caption{Running with network connectivity $p$ of percentage of estimate errors with respect to measurement errors for DC networks when $m=2N$ measurements are available for $N=20$ nodes with $<z>\sim 100$ and a sampling of 50 random networks and 100 measurements. \label{fig.stat_real_0_100_P}}
\end{center}
\end{figure}

With respect to the network connectivity we plot the dependence of the estimate errors as a function of $p$~(\ref{n_branch}) in figure~\ref{fig.stat_real_0_100_P} for the several functions $J$'s and the above choices.

\begin{figure}
\begin{center}
\includegraphics[width=130mm]{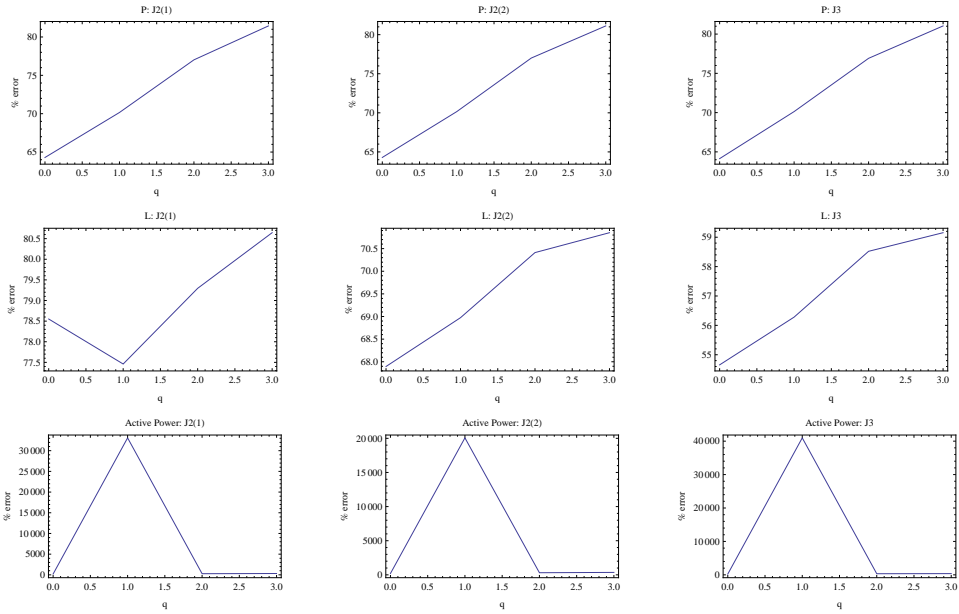}
\caption{Running with measurement availability $m=2N-q$ of percentage of estimate errors with respect to measurement errors for DC approximation to AC networks for $N=20$ nodes with $<z>\sim 100$ and $p=3$ and a sampling of 50 random networks and 100 measurements. \label{fig.stat_real_0_100_m}}
\end{center}
\end{figure}

\clearpage

\clearpage

\section{State Estimation in the presence of Gross Errors}

For a given network, once the analysis on the previous section is performed
such that a specific quadratic objective function and weight $\Sigma_i$ are chosen,
we can generally identify both gross measurement errors and network topological faults.

\begin{figure}
\begin{center}
\includegraphics[width=80mm]{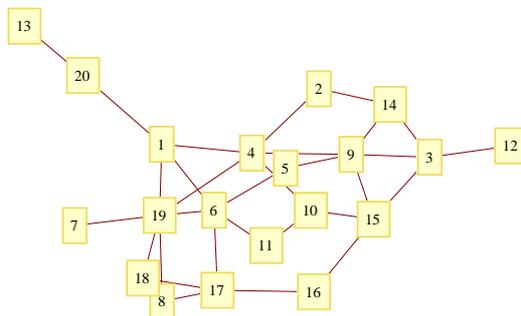}
\caption{Network topology considered for the simulation of the next figure~\ref{fig.03_gross_error_example}.\label{fig.03_network}}
\end{center}
\end{figure}

\begin{figure}
\begin{center}
\includegraphics[width=130mm]{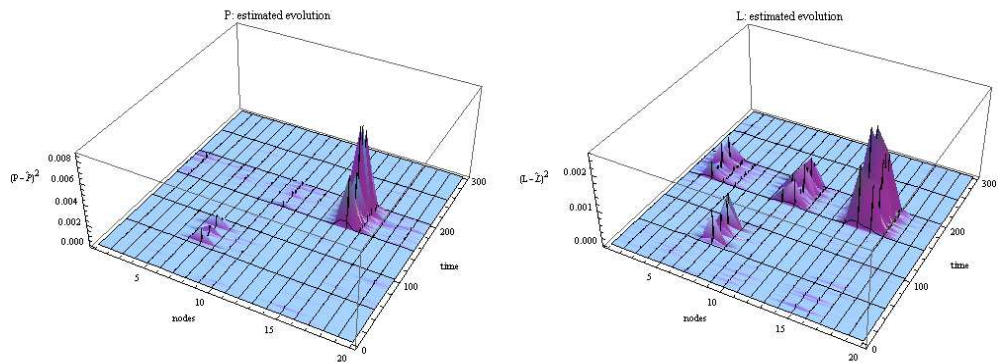}
\caption{Gross measurement errors effects on the estimation of $P$'s and $L's$: it is plotted the evolution, for each node, of $(P-\hat{P})^2$ and of $(L-\hat{L})^2$ for a DC approximation to an AC network with $N=20$ nodes, connectivity of $p=3$, standard measurement errors $\sigma_\%=0.01$
and gross errors of order $10\sigma_\%$. From time 50 there is a gross measurement error on the potential of node 7, then from time 100 this measurement is discarded such that the state estimation does not include the measurement $P_7$, then from time 150 there is gross measurement error on the injection of node 15, then from time 200 this measurement is also discarded such that the state estimation does not include neither the measurement $L_{15}$ neither $P_7$. Due to the gross measurement error of $L_{15}$ the estimates for the injections in adjacent nodes $\hat{L}_{3}$, $\hat{L}_{9}$ and $\hat{L}_{10}$ are also significantly affected.\label{fig.03_gross_error_example}}
\end{center}
\end{figure}

Typically, when a gross measurement error occurs at either a node potential or injection,
the deviations from the Kirchoff laws become more significant at that node such that
the difference between the measured quantities and estimated quantities become much larger
than in the absence of gross measurement errors. Identifying such discrepancies allows to
identify the measurements containing these errors and discard them when estimating the network
state. We exemplify the occurrence and discarding of such gross measurement errors for the network
represented in figure~\ref{fig.03_network} in
figure~\ref{fig.03_gross_error_example}.  

In addition we note that often, when a gross measurement error occurs at a given node, depending on the specific
network topology, it may affect significantly the estimates for the quantities in adjacent nodes. This occurrence is
also exemplified in figure~\ref{fig.03_gross_error_example}.

To detect the existence of such gross measurement errors it is enough to set a threshold for the quantities $(P_i-\hat{P}_i)^2$ and $(L_i-\hat{L}_i)^2$ above which a correction procedure is trigged checking whether the value of the quadratic objective function is lower when the specific nodal measurements are discarded. If this is the case, the measurements are discarded. A refinement of this procedure may include the checking of the neighbors nodal measurements as well as next neighbors nodal measurements. Such refinement will make the detection procedure
slower. We also note that it is required that the ration between a gross error and the standard white noise level be significant,
otherwise these two sources of measurement error are not distinguishable. 

\clearpage
\section{State Estimation in the presence of topological faults}

A topological fault constitutes an unaccounted opening or closing of a switch $s_{ik}$ such
that the respective branch $ik$ is erroneously represented in the matrix $Y_{\mathrm{state}}$
employed in the definition of the functions $J$ and state estimation. Hence, for a given set of
potentials $P$ and injections $I$, mathematically the problem of topology estimation can be formulated
as the integer NP-hard problem of estimating the quantities defining the switchers state $\hat{s}_{ik}$
\begin{equation}
\left(Y_{on}-\sum_{ik}\frac{\hat{s}_{ik}}{z_{ik}}\,M_{ik}\right)P=L\ \ ,\ \ \hat{s}_{ik}\in\left\{0,1\right\}\ ,
\end{equation}
where $Y_{on}$ corresponds to the network admittance matrix with all the switchers on and the
sum is over the branches where switchers are present and the matrices $M_{ik}$ generally constitute an
orthonormal basis for the admittance matrix representing the contribution of each branch, i.e. the
matrix entries $ii$ and $kk$ are $1$ and the entries $ik$ and $ki$ are $-1$ 
\begin{equation}
M_{ik}=\begin{array}{l}\ \ \ \ \ \ \ \ \ \ \ \ \ \ \ \ \ \ \ \ \ (i)\ \ \ \ \ \ \ \ \ (k)\\\begin{array}{c} \\ \\ (i)\\ \\ (k)\\ \\ \\\end{array}\left[\begin{array}{ccccccc}
0&\cdots&0&\cdots&0&\cdots&0\\
\vdots& &\vdots& &\vdots& &\vdots\\
0&\cdots&+1&\cdots&-1&\cdots&0\\
\vdots& &\vdots& &\vdots& &\vdots\\
0&\cdots&-1&\cdots&+1&\cdots&0\\
\vdots& &\vdots& &\vdots& &\vdots\\
0&\cdots&0&\cdots&0&\cdots&0
\end{array}\right]\end{array}
\end{equation}
Specifically this problem can be solved by employing either heuristic algorithms (e.g. Simplex), discrete
numerical methods (e.g. Gradient method) or enumeration.

When only a subset of the topology is unknown the problem is significantly simplified.
In particular if a single switch state $ik$ is unknown we obtain that
\begin{equation}
\begin{array}{rcl}
M_{ik}P&=&\left[\begin{array}{ccccccc}
0&\cdots&0&\cdots&0&\cdots&0\\
\vdots& &\vdots& &\vdots& &\vdots\\
0&\cdots&+1&\cdots&-1&\cdots&0\\
\vdots& &\vdots& &\vdots& &\vdots\\
0&\cdots&-1&\cdots&+1&\cdots&0\\
\vdots& &\vdots& &\vdots& &\vdots\\
0&\cdots&0&\cdots&0&\cdots&0
\end{array}\right]\left[\begin{array}{c}P_1\\\vdots\\P_i\\\vdots\\P_k\\\vdots\\P_N\end{array}\right]\\[25mm]
&=&\left[\begin{array}{c}0\\\vdots\\(P_i-P_k)/z_{ik}\\\vdots\\(P_k-P_i)/z_{ik}\\\vdots\\0\end{array}\right]
=\left[\begin{array}{c}0\\\vdots\\F_{ik}\\\vdots\\-F_{ik}\\\vdots\\0\end{array}\right]
\end{array}
\end{equation}

\begin{figure}
\begin{center}
\includegraphics[width=80mm]{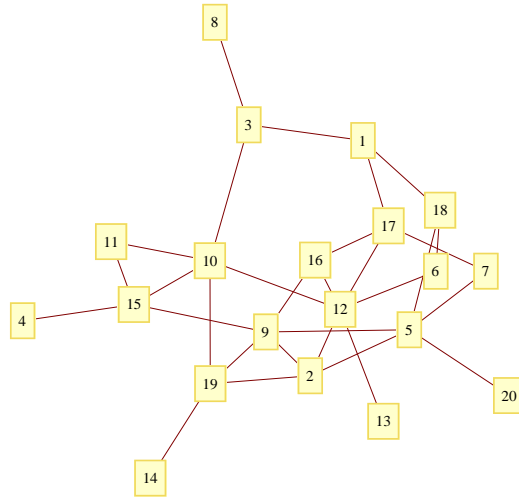}
\caption{Network topology considered for the simulation of the next figure~\ref{fig.04_fault_correction_example}.\label{fig.04_network}}
\end{center}
\end{figure}

\begin{figure}
\begin{center}
\includegraphics[width=130mm]{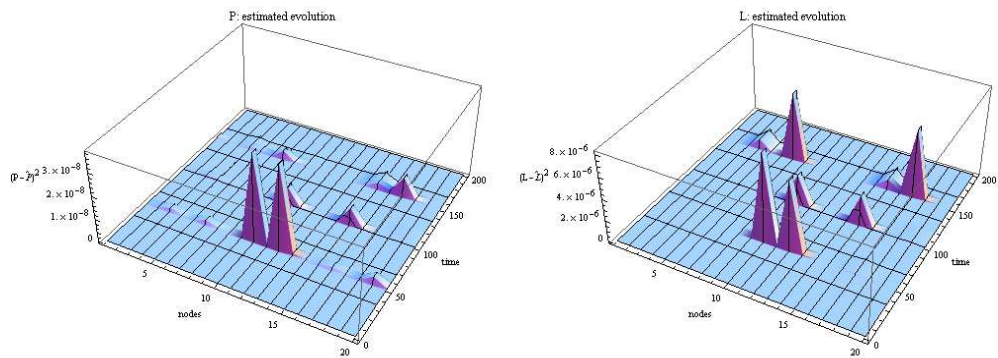}
\caption{Fault correction based on the estimation of $P$'s and $L's$: it is plotted the evolution, for each node, of $(P-\hat{P})^2$ and of $(L-\hat{L})^2$ for a DC approximation to an AC network with $N=20$ nodes, connectivity of $p=3$, standard measurement errors $\sigma_\%=0.05$
and simultaneously topology faults of branch $(12,17)$ and $(10,12)$ at time 50, of branch $(10,15)$ and $(9,19)$ at time 100 and of branch $(4,15)$ and $(7,17)$ at time 150. When the fault is detected, the assumed topology for state estimation is corrected such that the quantities $(P-\hat{P})^2$ and $(L-\hat{L})^2$ return to the standard value corresponding to the measurement white noise.\label{fig.04_fault_correction_example}}
\end{center}
\end{figure}

For a given state estimation for the $P$'s and $L$'s, this result allows to identify whether the assumed topology is correctly estimated or not.
When a wrong topology is assumed for the switcher $ik$ the quantities $(P_i-\hat{P}_i)^2$, $(P_k-\hat{P}_k)^2$, $(L_i-\hat{L}_i)^2$ and $(L_k-\hat{L}_k)^2$ are much bigger than the ones for the remaining nodes. Setting a threshold for these quantities generally allows to identify at least one of the nodes $i$ or $k$. Once a faulty node is identified, flipping the switchers connecting to the identified node and comparing the quadratic objective function for the alternative topologies obtained it is chosen the topology that minimizes the function $J$ being employed such
that the assumed topology os corrected. This procedure requires only as many computations as the closest integer to $p$ (the network connectivity).
For the network with topology represented in figure~\ref{fig.04_network} we exemplify these procedure for two simultaneously faults in figure~\ref{fig.04_fault_correction_example}. Although correcting only one fault at a time it successfully corrects several faults successively.

\clearpage
\section{Generic State Estimation}

Let us now describe how to implement a procedure to fault detection
and correction in a generic network with unknown topology and state.
From the previous section we have concluded that we required:
\begin{itemize}
\item to estimate the nodal potentials $\hat{P}$ and injections $\hat{L}$;
\item to estimate full topology, hence defining a network initial state;
\item to estimate the network topology evolution.
\end{itemize}
Hence to actually implement such a procedure we are considering two distinct steps which should run cyclically:
\begin{itemize}
\item globally estimate full topology by an integer programming algorithm;
\item locally estimate and correct topological faults and gross errors as network evolves.
\end{itemize}
The first step is slow, however must be run periodically to reset the network
to a known reliable state. The second stage is faster, however the faults are inspected locally,
hence is not as reliable as the first step. We consider the following computational method:
\begin{enumerate}
\item estimate the potentials $\hat{P}$ and injections $\hat{L}$ by minimizing a given quadratic objective function $J$ chosen from the ones discussed previously;

\item estimate the full topology defining the initial state by solving the full integer NP-Hard problem, i.e. find a feasible solution for
the set of linear constraints
\begin{equation}
	(Y\hat{P} – \hat{L}) - \epsilon \leq \sum_{ik}\frac{s_{ik}}{z_{ik}} M_{ik} \hat{P} \leq (Y\hat{P} – \hat{L}) + \epsilon\ \ ,\ \ s_{ik} \in \{0,1\}\ ,
\end{equation}
where $\epsilon$ is estimated from the errors for $\hat{P}$ and $\hat{L}$. 

\item estimate the network topology evolution by identifying and correcting the topology faults and gross errors:
\begin{enumerate}	

\item set a noise level threshold $\delta$ for the potentials $\delta_P$, injections $\delta_L$ and optionally to the active power $\delta_{ap}$ and/or reactive power $\delta_{rp}$;

\item detect the possibility of fault by defining at each node $k$ the forward and backward average of $n$ measurements estimates $<P_k-\hat{P}_k>_b$, $<P_k-\hat{P}_k>_f$, $<L_k-\hat{L}_k>_b$, $<L_k-\hat{L}_k>_f$, $<pa_k>_b$, $<pa_k-\hat{pa}_k>_f$, $<pr_k-\hat{pr}_k>_b$ and $<pr_k-\hat{pr}_k>_f$, where the index $b$ stands for 'backward' and the index $f$ stands for 'forward'. At each node identify if the fault may exist by checking the following conditions
\begin{equation}
\begin{array}{rcl}
\displaystyle\frac{<P_k-\hat{P}_k>_f}{<P_k-\hat{P}_k>_b}&>&\delta_P\ \ \mathrm{or}\\[5mm]
\displaystyle\frac{<L_k-\hat{L}_k>_f}{<L_k-\hat{L}_k>_b}&>&\delta_L\ \ \mathrm{or}\\[5mm]
\displaystyle\frac{<ap_k-\hat{ap}_k>_f}{<ap_k-\hat{ap}_k>_b}&>&\delta_{ap}\ \ \mathrm{or}\\[5mm]
\displaystyle\frac{<rp_k-\hat{rp}_k>_f}{<rp_k-\hat{rp}_k>_b}&>&\delta_{rp}\\[5mm]
&&\displaystyle\  \ \Rightarrow\ \ \mathrm{fault\  may\  exist}\ .
\end{array}
\end{equation}
\item in the possibility of the existence of a fault, identify if it is actually a fault and, if it is, correct it
\begin{itemize}
\item select the 2 nodes $k_1$ and $k_2$ corresponding to the higher error for the quantity that trigged the possibility of a fault;
\item record $min_0 = J$ for assumed known topology prior to the detection of possibility of fault;
\item flip independently each of the switchers $(k_1,i_1)$ and $(k_2,i_2)$ adjacent to nodes $k_1$ and $k_2$ and compute $min_{k_1i_1}=J$ and $min_{k_2i_2}=J$ for each topology corresponding to the flip of the several switchers. The lower value of the evaluated functions $min$ corresponds to the best topology;
\item hence if exists a $min_{ki}$ lower than $min_0$ the fault is identified and the assumed known topology can be updated;
\item if the lower value for the functions is $min_0$ there is no fault and the assumed known topology is not updated. If this is the case record 	  the nodes $k_1$ and $k_2$ as possible sources of gross measurement error.
\end{itemize}

\item identify the existence of gross errors. If same node k is often recorded as a source of gross errors more than some predefined number of times $Ng$, remove the measurements for node k and eventually send a maintenance team to check the measurement instrumentation for this node. 
\end{enumerate}
\end{enumerate}
We note that the evolution algorithm clearly distinguish the existence of gross errors from topological faults. When a gross error
occurs the flipping of the switchers does not decrease the function $J$.

\begin{figure}
\begin{center}
\includegraphics[width=70mm]{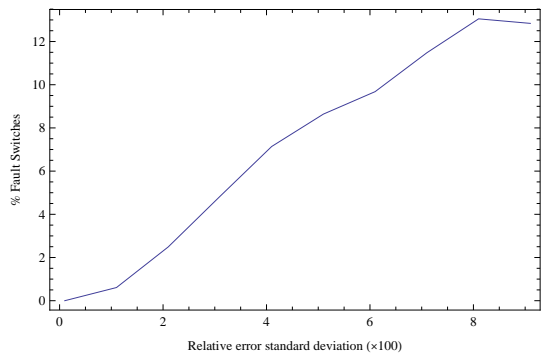}
\caption{The percentage of switchers states wrongly estimated for a global estimation of the network topology as a function of the measurement error $\sigma_\%$. \label{fig.stat_0_faults}}
\end{center}
\end{figure}

\begin{figure}
\begin{center}
\includegraphics[width=120mm]{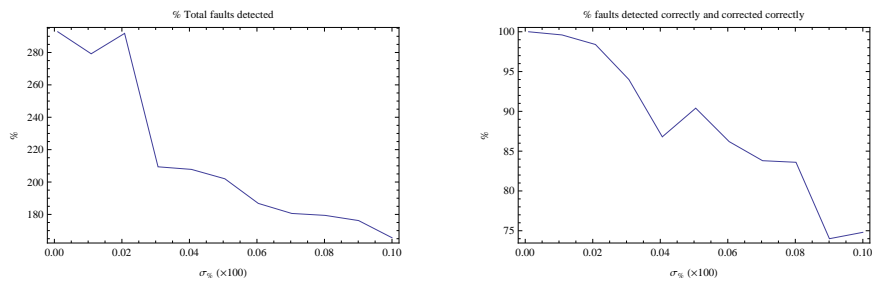}
\caption{The percentage of faults correctly corrected and the number of identified and analyzed possible faults as a function of $\sigma_\%$ on DC approximation to AC networks of $N=20$ nodes considering 5 random topologies and 50 random events of 2 consecutive faults. \label{fig.stat_1_faults}}
\end{center}
\end{figure}

The efficacy and efficiency of the several stages of this method significantly depends on the value of $\sigma_\%$.
In figure~\ref{fig.stat_0_faults} it is plotted the rising of the percentage of switcher states wrongly estimated as a function of $\sigma_\%$
when the global network state is estimated. In~\ref{fig.stat_1_faults} it is plotted the evolution of the percentage of switchers faults correctly
corrected and the percentage of adjacent switchers analyzed. As it is readily verified only for relatively low $\sigma_\%<0.01$ the method has a success of fault detection and correction over $99\%$.

\clearpage
\section{Fault detectability}

\begin{figure}
\begin{center}
\includegraphics[width=50mm]{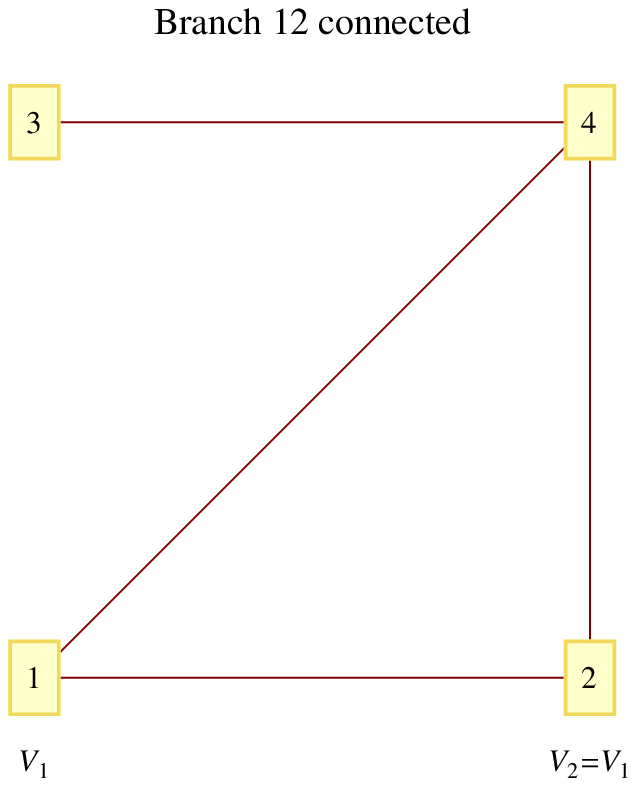}\ \ \ \ \includegraphics[width=50mm]{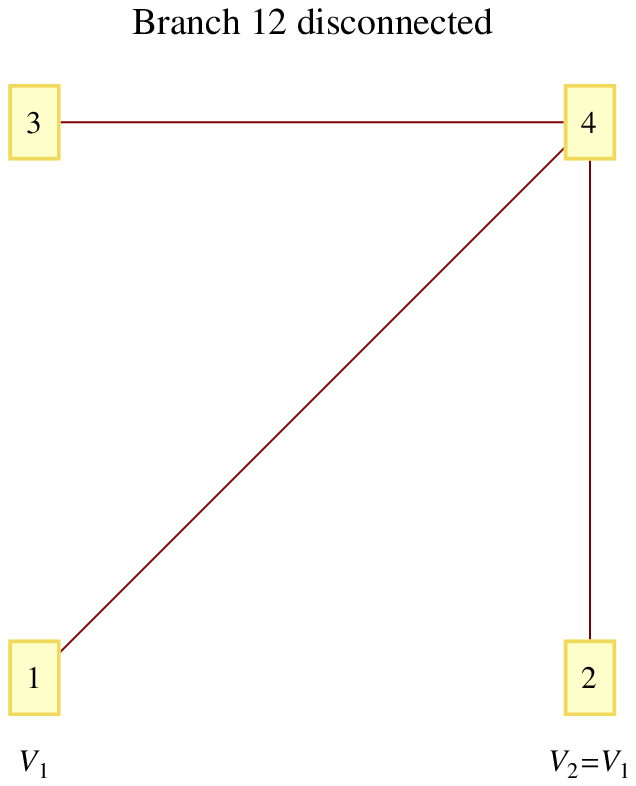}
\caption{Example of distinct indistinguishable networks. As $V_1=V_2$ there is no flux in the branch $12$, $F_{12}=0$. \label{fig.degeneracy_example}}
\end{center}
\end{figure}

Given exact values for potentials $P_0$ and injections $L_0$ and two distinct topologies allowing for this network state $Y_1$ and $Y_2$ we obtain the linear system: 
\begin{equation}
\left\{\begin{array}{rcl}Y_1 P_0&=&L_0\\[5mm]Y_2 P_0&=&L_0\end{array}\right.\ \ \Leftrightarrow\ \ (Y_1-Y_2)P_0=0\ .
\end{equation}
Reversely, given two distinct topologies, these are mathematically indistinguishable for every exact value of $P_0$ which is a solution of these equations. The solutions to this system of equations correspond to the null space of the matrix $Y_1-Y_2$. Considering the matrix basis $M_{ik}$
it is straight forward to obtain the solution:
\begin{equation}
M_{ik} P = 0\ \ \Leftrightarrow\  \ P_i = P_k\ ,
\end{equation}
such that the topologies differing by the flip of the switch $ik$ with $P_i = P_k$ are not distinguishable, although both topologies
may be admissible in a real network. The example of two such topologies, differing only by the state of the switch $12$ is pictured in figure~\ref{fig.degeneracy_example}.

However with measurement white noise
\begin{equation}
P_i + \epsilon_{P,i} = P_k + \epsilon_{P,k}\ .
\end{equation}
is a possible physical condition, hence measurements with a relatively small projection in the orthogonal space to the null space of the difference of admittance matrices corresponding to distinct topologies are mathematically indistinguishable.

\begin{figure}
\begin{center}
\includegraphics[width=120mm]{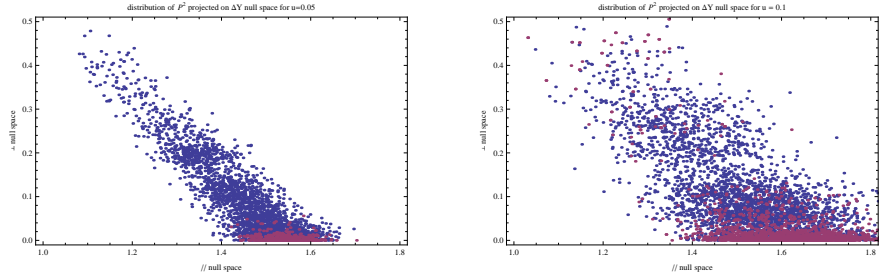}
\caption{Representation of the measured potentials vector $P^2=\sum P_i^2$ in the null space of the matrix $\Delta Y=Y_{exact}-Y_{known}$
for DC approximation to AC networks of $N=20$ nodes considering 5 random topologies and 50 random events of 2 consecutive faults for
$\sigma_\%=0.05$ and $\sigma_\%=0.1$. Blue: faults corrected correctly; Magenta: faults corrected wrongly. \label{fig.stat_degeneracy}}
\end{center}
\end{figure}

We verify that this is the main cause for the undetectability of faults. Considering a statistical sample of several distinct topologies and
measurement errors the faults which are not detectable, hence not corrected by the method described in the previous section, correspond to
measurements for which the potentials vector is nearly parallel to the null space of the matrix $\Delta Y = Y_{exact}-Y_{known}$, i.e. the
change of topology due to the fault being detected. This result is plotted in figure~\ref{fig.stat_degeneracy}. 

\begin{figure}
\begin{center}
\includegraphics[width=120mm]{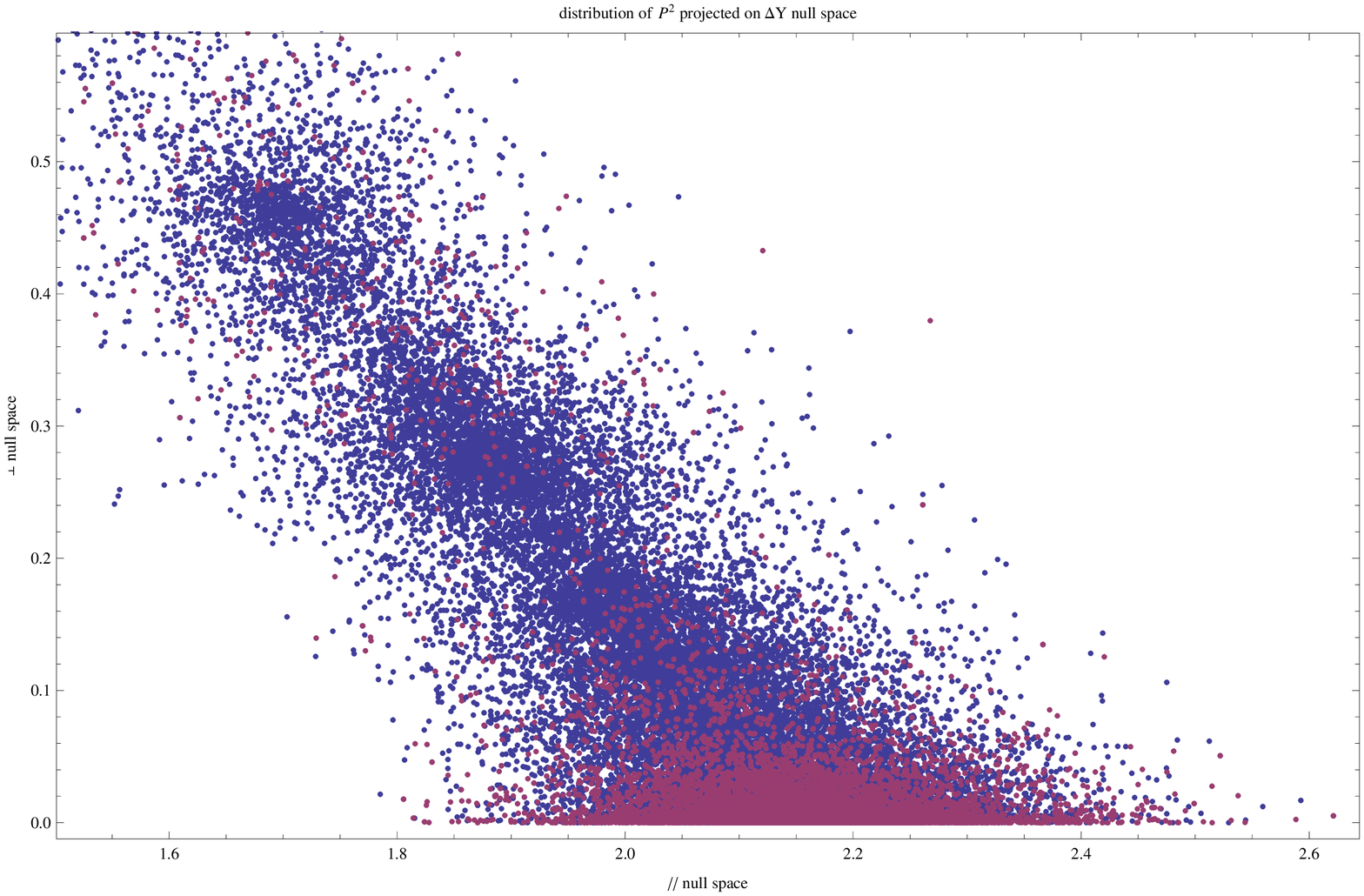}
\caption{Representation of the measured potentials vector $P^2=\sum P_i^2$ in the null space of the matrix $\Delta Y=Y_{exact}-Y_{known}$
for DC approximation to AC networks of $N=20$ nodes considering 5 random topologies and 50 random events of 2 consecutive faults for
$\sigma_\%\in]0,0.1]$. Blue: faults corrected correctly; Magenta: faults corrected wrongly. \label{fig.stat_degeneracy_2}}
\end{center}
\end{figure}

\begin{figure}
\begin{center}
\includegraphics[width=120mm]{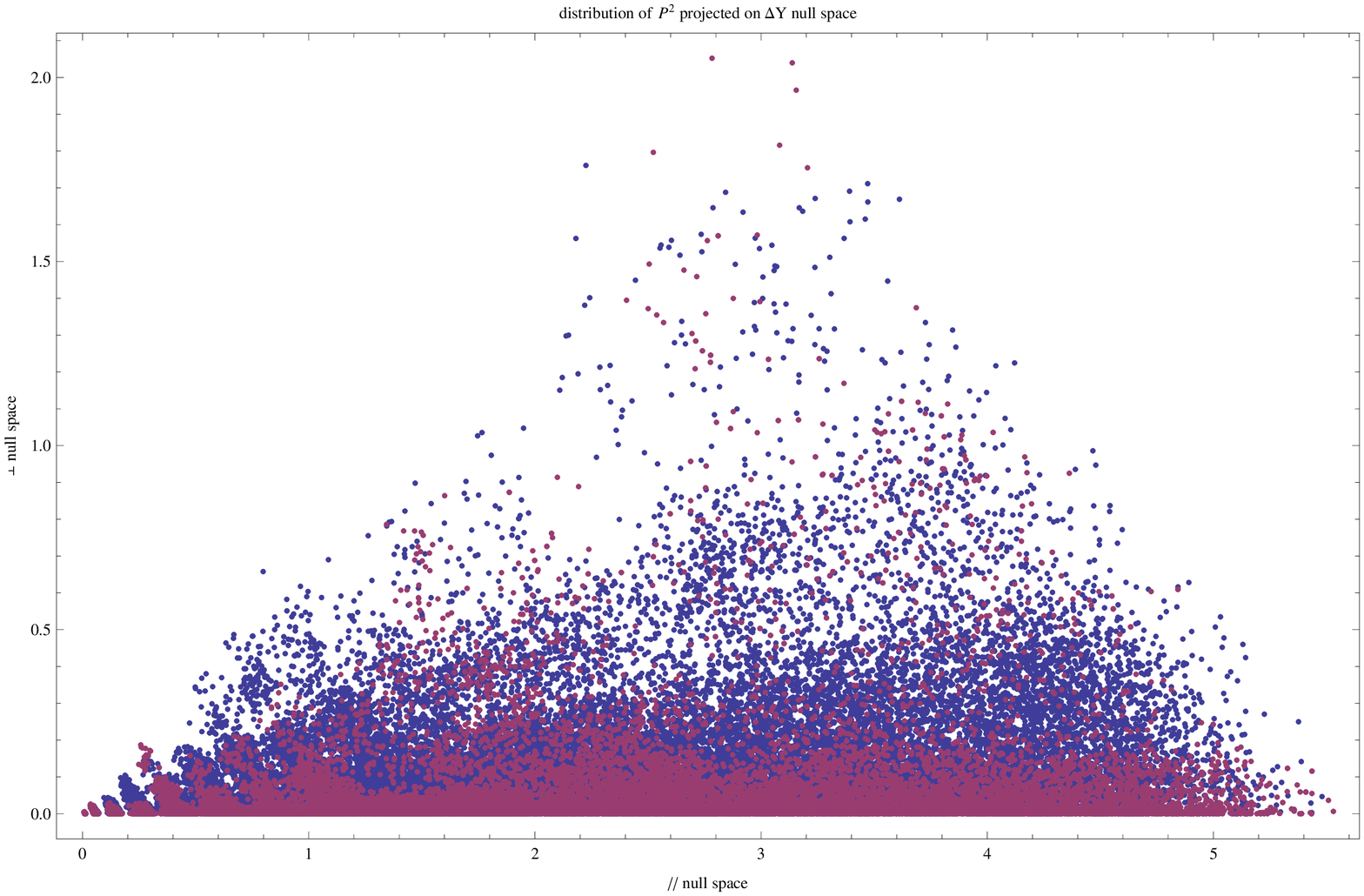}
\caption{Representation of the measured potentials vector $P^2=\sum P_i^2$ in the null space of the matrix $\Delta Y=Y_{exact}-Y_{known}$
for DC approximation to AC networks of $N=20$ nodes considering 5 random topologies and 50 random events of 2 consecutive faults for
$\sigma_\%=0.1$ for a sample of 15 distinct values for the average value of the node potentials $\left<P_i\right>\in]0,2]$. Blue: faults corrected correctly; Magenta: faults corrected wrongly. \label{fig.stat_degeneracy_3}}
\end{center}
\end{figure}

\section{Conclusions and recommendations}

Hence we have fully described an algorithm to estimate power network state. We have concluded that the main source of uncertainty is
the existence of indistinguishable topologies. This is a well known problem~\cite{pnetworks_1,pnetworks_2} being also the main mechanism
that allows for successful attacks in communication networks~\cite{security}.

The particular algorithm described here requires to fine-tune the quadratic objective function as well as the remaining parameters (weights, noise threshold, etc). For specific networks with an higher number of nodes it is required to redo the analysis and statistics carried in this report
to optimize the detection and correction of topology and measurement errors. We further note that, when aiming at estimating the values of the power fluxes, instead of the nodal potentials and injections, it is required an explicit dependence of these quantities in the quadratic objective function. As possible detectability improvement it may be considered the checking of next-neighbors nodes and/or switchers, however the method is slower.\\[5mm]

\noindent{\bf Acknowledgments}\\[3mm]
\noindent Work developed within the scope of the strategical project of GFM-UL PEst-OE/MAT/UI0208/2011.
PCF work supported by FCT-MCTES grant SFRH/BPD/34566/2007.

\end{document}